\documentclass[useamsfonts,twocolumns]{pasj00}
\draft

\begin{document}
\SetRunningHead{Y. Moritani et al.}{Letter:Optical Spectroscopy of A0535+262 during the giant outburst}
\Received{xxxx/xx/xx}
\Accepted{xxxx/xx/xx}

\title{
Drastic Spectroscopic Variability of the Be/X-ray Binary A0535+262/V725 Tau 
during and after the 2009 Giant Outburst}

\author{Yuuki \textsc{Moritani},\altaffilmark{1}
			Daisaku \textsc{Nogami},\altaffilmark{2}
			Atsuo T. \textsc{Okazaki},\altaffilmark{3}
			Akira \textsc{Imada},\altaffilmark{4}
			Eiji \textsc{Kambe},\altaffilmark{4}\\
			Satoshi \textsc{Honda},\altaffilmark{2}
			Osamu \textsc{Hashimoto},\altaffilmark{5} and 
			Kohei \textsc{Ichikawa}\altaffilmark{1}} %
\altaffiltext{1}{Department of Astronomy, Faculty of Science, Kyoto University, Sakyo-ku, Kyoto 606-8502}
\altaffiltext{2}{Kwasan Observatory, Kyoto University, Yamashina-ku, Kyoto 607-8471}
\altaffiltext{3}{Faculty of Engineering, Hokkai-Gakuen University, Toyohira-ku, Sapporo 062-8605}
\altaffiltext{4}{Okayama Astrophysical Observatory, National Astronomical Observatory of Japan, Asakuchi, Okayama 719-0232}
\altaffiltext{5}{Gunma Astronomical Observatory, Takayama-mura, Gunma 377-0702}
\email{moritani@kusastro.kyoto-u.ac.jp}


\maketitle

\begin{abstract}
We report on high-dispersion optical spectroscopic observations of the Be/X-ray binary A0535+262/V725 Tau during the giant outburst in November/December 2009 and after it.
The observed emission line profiles, reflecting the structure of the geometrically thin circumstellar envelope of the Be star (Be disk), show drastic variabilities and indicate the existence of a warped component.
The enhanced blue shoulder seen after periastron passage implies the gas stream from a dense part of the Be disk to the neutron star.
\end{abstract}

\section{Introduction}
Be/X-ray binaries, which are one of the major subgroups of high-mass X-ray binaries, along with superfast X-ray transients and supergiant/X-ray binaries, consist of a Be star and a compact object, a neutron star in general.
Be stars are B-type giant or dwarf stars (luminosity class III -- V) which have exhibited Balmer lines in emission at least once.
They have a geometrically thin circumstellar envelope called a Be disk, formed as a result of viscous diffusion ($\lesssim$ 1 km/s) of matter ejected from an equatorial region of a rapidly rotating central star [\citet{Lee91}; see also a recent review by \citet{Car10}].

The activity of Be/X-ray binaries in X-rays is divided into three categories with respect to the luminosity: quiescent state ($L_X \lesssim 10^{36}$ erg s$^{-1}$), normal (type I) outbursts ($L_X \sim 10^{36-37}$ erg s$^{-1}$), and giant (type II) outbursts ($L_X \gtrsim 10^{37}$ erg s$^{-1}$).
The normal outbursts occur around periastron passage, and last for several days.
This type of outbursts are seen in systems with intermediate to high eccentricities where the mass transfer from the Be disk to the neutron star takes place at every periastron passage \citep{Oka01b,Neg01a}.
The giant outbursts, on the other hand, lasting several tens of days, are not well understood since they show no orbital modulation and occur much less frequently than normal outbursts.

A0535+262/V725 Tau is one of the best studied Be/X-ray binaries since its discovery in 1975 \citep{Ros75, Coe75}.
This system consists of an X-ray pulsar orbiting around an O9.7IIIe star \citep{Gia80} in a relatively wide ($P_{\rm orb}$ $\sim$ $110$ d) and eccentric ($\sim$ $0.47$) orbit \citep{Fin94}.

The giant outbursts of A0535+262 have been observed eight times, in 1975, 1977, 1980, 1989, 1994, 2005, 2009 and 2011 [see \citet{Cam11b} and references theirin].
Based on long-term photometric observations including the giant outburst in 1994, \citet{Hai04} suggested that the reduction of the truncation radius (mass transfer) commenced before the giant outburst.
For other systems, several observations including giant outbursts have been reported: EXO 2030+375 \citep{Wil02,Wil08}, 4U0115+63 \citep{Neg01b, Rei07}, for instance.
\citet{Wil08} found that the orbital phase of the normal outburst around a giant one is shifted by the excitation of a density wave.
\citet{Neg01b} proposed that enhanced, broad H$\alpha$ line profiles observed around each giant outburst are from a precessing warped disk.

In November/December 2009, a giant outburst occurred in A0535+262.
It started brightening in X-rays on 30 November, with a couple of precursors in October \citep{Sug09}.
The Swift/BAT team reported that the peak of the giant outburst reached 3.1 Crab in 15 - 50 keV.
After the giant outburst, normal outbursts occurred around the three consecutive periastron passages; in March, July and October 2010, followed by the next giant outburst in February 2011.

In this letter, we report on the results from the optical high-dispersion spectroscopic observations during and after the 2009 giant outburst.
This is the first-ever high-dispersion monitoring observations that cover the whole period of a giant outburst.
The detailed analysis of the whole dataset will be published in a forthcoming paper.

\section{Observations}
Optical spectroscopic observations of A0535+262 were carried out on 35 nights from November 2009 to September 2010 at the Okayama Astrophysical Observatory (OAO) with a 188 cm telescope equipped with HIDES (High Dispersion Echelle Spectrograph), and at Gunma Astronomical Observatory (GAO) with a 1.5 m telescope equipped with GAOES (Gunma Astronomical Observatory Echelle Spectrograph).
Since December 2009, most of the HIDES spectra were obtained with its new fiber-feed system.
HIDES covers 3500 -- 6800 \AA \ range with three 2k $\times$ 4k EEV 42 -- 80 CCDs.
The wavelength coverage of GAOES is 4800 -- 6700 \AA \ and the detector is a EEV 44 -- 82 CCD.
The typical wavelength resolution $R$ and the signal to noise ratio $S/N$ of our HIDES data around H$\alpha$ are $R$ $\sim$ 60000 and  $S/N$ $\gtrsim$ 100, respectively.
On the other hand, our H$\alpha$ data obtained with GAOES has $R$ $\sim$ 30000 and $S/N$ $\sim$ 100.
The effective exposure time was 3600 to 5400 seconds with HIDES, and 3600 to 9600 seconds with GAOES, respectively.

The obtained data were reduced in the standard way, using IRAF\footnote{http://iraf.noao.edu/} echelle package -- subtraction of bias, flat fielding, calibration of the wavelength using Th-Ar lines, normalization of the continuum, and helio-centric correction of the radial velocity.

\begin{figure}
  \begin{center}
    \includegraphics[width=222pt]{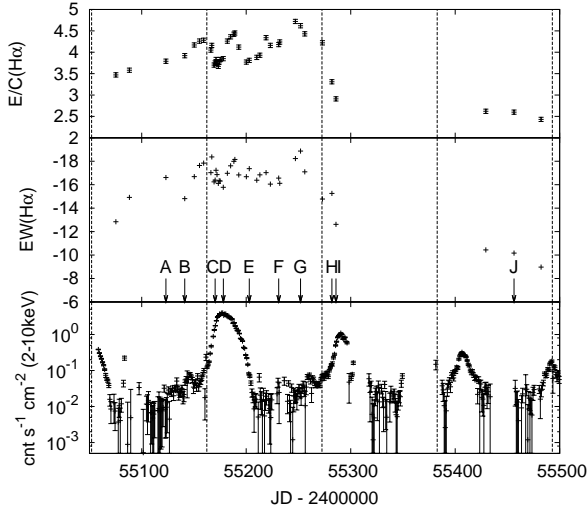}
  \end{center}
  \caption{Intensity of the normalized peak E/C ({\it top}) and equivalent width EW ({\it middle}) of the obtained H$\alpha$ line profiles. 
{\it Bottom}: X-ray light curve in 2 - 10 keV by MAXI/GSC. 
The dotted lines indicate the $\phi_X$ = 0 in \cite{Mor10}.
The short arrows in the middle panel denote the HJD of 11 representative spectra displayed in figure \ref{Spectra_rep}.
}
  \label{LC_X}
\end{figure}

Figure \ref{LC_X} shows the normalized peak intensity and the equivalent width of the H$\alpha$ line from these observations.
Figure \ref{LC_X} also shows the X-ray light curve of MAXI/GSC\footnote{http://maxi.riken.jp/top/index.php?cid=1\&jname=J0538+263} (2 -10 keV) for comparison.
The vertical dotted lines indicate the rising time of the normal outburst, $\phi_X$ = 0, estimated by Moritani et al. (2010) (hereafter, \authorcite{Mor10}).
Our observations cover the giant outburst in November/December 2009, in the rising phase more densely than in the fading phase.
After the giant outburst, several monitoring observations were carried out until the next periastron passage in March 2010, when a normal outburst occurred, followed by three more observations about half a year later.

\section{Results}
\begin{figure*}
  	\begin{minipage}{0.5\hsize}
  \begin{center}
    \includegraphics[width=174pt]{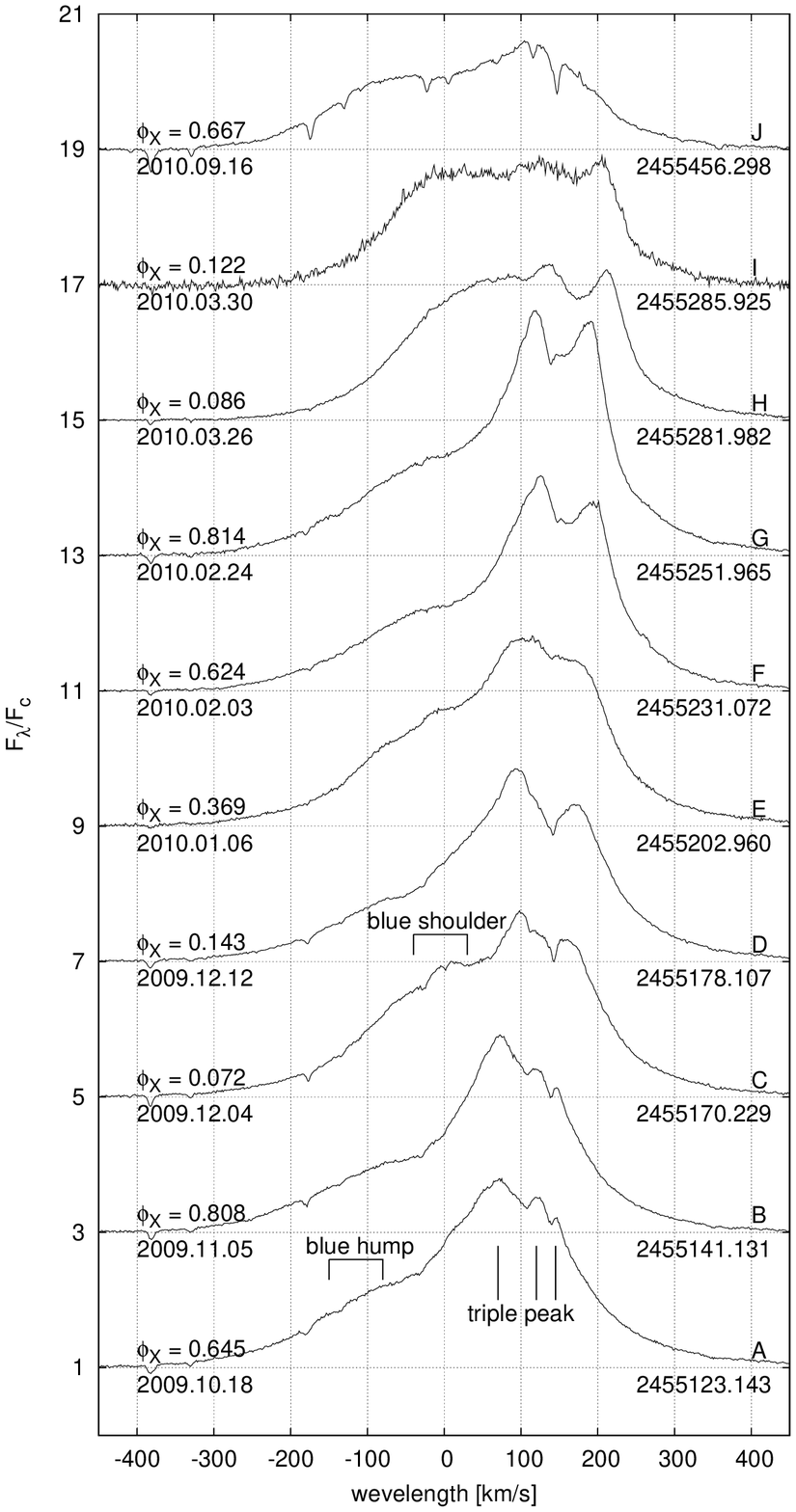}
  \end{center}
  	\end{minipage}
  	\begin{minipage}{0.5\hsize}
  \begin{center}
    \includegraphics[width=174pt]{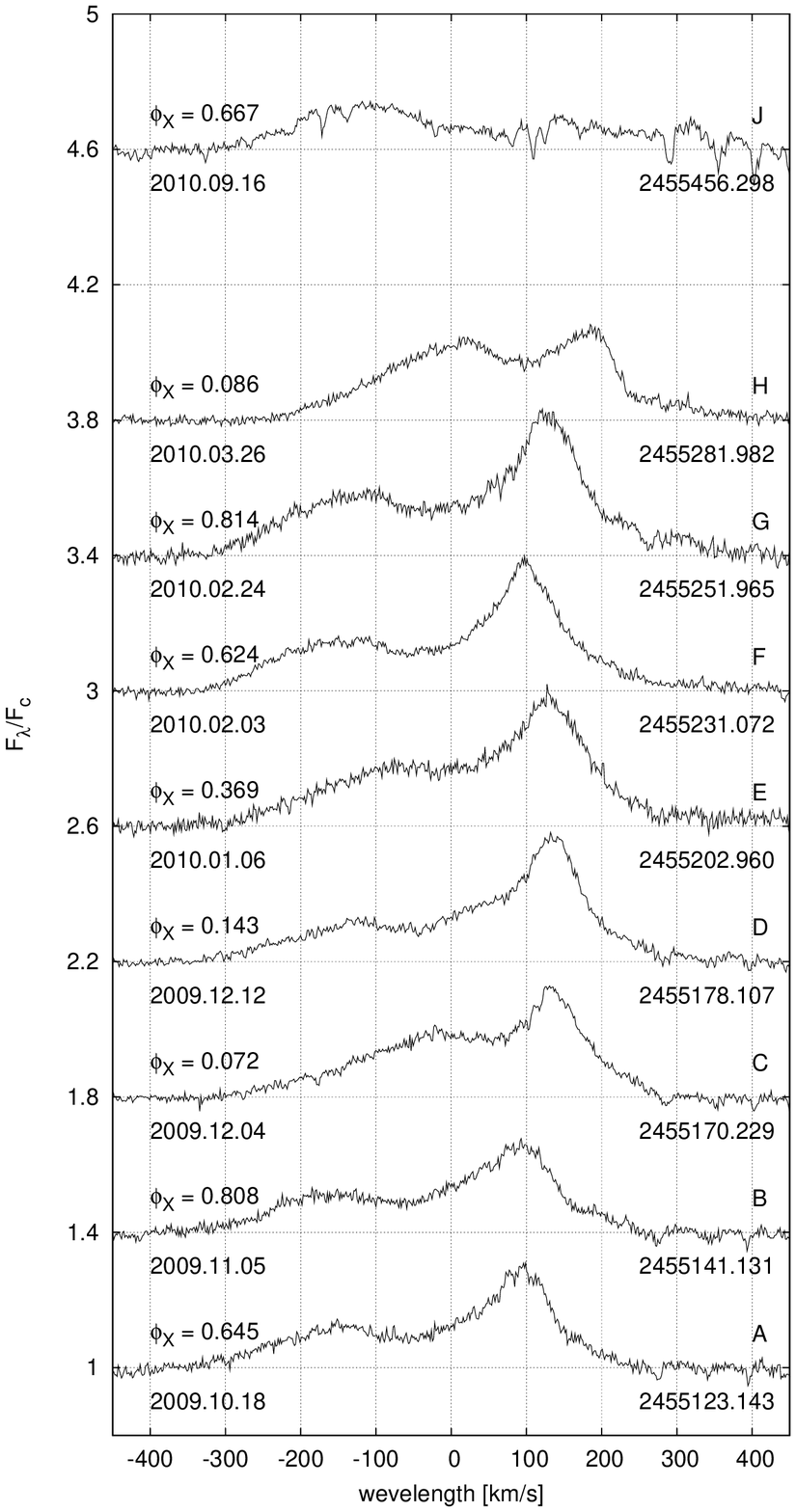}
  \end{center}
  	\end{minipage}
  \caption{Representative spectra of H$\alpha$ and He I $\lambda$5876 lines.
  The observation date and HJD of mid exposure time are written near the profile on the left and right side, respectively.
  The profiles showed drastic variability (see text for details).
}
  \label{Spectra_rep}
\end{figure*}

The representative H$\alpha$ and He I $\lambda$5876 profiles from our observations are respectively shown in the left and right panels in figure \ref{Spectra_rep}. 
For reasons of clarity, the spectra have linear offsets along the vertical axis from each other. 
The observation date and HJD of the mid exposure time are annotated below each profile on the left and right sides, respectively.
The phase $\phi_X$ estimated by \cite{Mor10} is also shown above each profile. 
He I $\lambda$5876 line profile on 30 March 2010 is not presented because of its poor S/N.

Figure \ref{Spectra_rep} shows that the H$\alpha$ line profile has drastically changed during the giant outburst (spectra A-E) and after it (spectra F-J). 
The H$\alpha$ line profile during this period is characterized by a strongly redshifted triple peak and a broad hump/shoulder in the blue wing.
The variation of the He I $\lambda$5876 line profile was also significant, although it was not so drastic as that of the H$\alpha$ line.
The He I $\lambda$5876 line profile was more or less a typical double peaked profile seen in many Be stars.

In October 2009, before the giant outburst, the triple peak of the H$\alpha$ line was already strongly redshifted, while the He I $\lambda$5876 line profile was double peaked with V $\ll$ R (spectrum A).
In a month, during the precursor of the giant outburst (\cite{Wil09}; see also figure \ref{LC_X}), the intensity of normalized peak, E/C, of both lines significantly increased, and the blue hump superposed on the smooth slope of the H$\alpha$ line brightened (spectrum B, see also top panel of figure \ref{LC_X}). 
The blue component of the He I $\lambda$5876 line profile also became brighter.

Two days before the peak of the giant outburst (6 December 2009, \cite{Kri09}), the H$\alpha$ line profile changed drastically; the E/C ratio decreased and a bright ``shoulder'' appeared around $0\;{\rm km\;s}^{-1}$ in the blue wing (spectrum C). 
A similar, bright feature was seen in the central part of the He I $\lambda$5876 line profile  (between $-50\;{\rm km\;s}^{-1}$ and $+50\;{\rm km\;s}^{-1}$).
However, these features were gone by 12 December 2009, six days after the X-ray maximum (spectrum D).

Three weeks later (6 January 2010), in the fading phase of the giant outburst, the blue shoulder in the H$\alpha$ line reappeared (spectrum E). 
The E/C ratio started decreasing by then, after it gradually increased until 22 December 2009. 
The He I $\lambda$5876 line showed similar variability; the E/C ratio increased and then slightly decreased again.

In February 2010, the double peak of the H$\alpha$ line continued to grow (spectra F and G) and reached the highest intensity during the precursor of a normal outburst  (24 February 2010, spectrum G). 

On 26 March 2010, near the peak of the normal outburst in the orbital cycle next to the giant outburst, the profile drastically changed; in the H$\alpha$ line, the strongly redshifted double-peak weakened and a huge, broad shoulder appeared in the blue wing, while in the He I $\lambda$5876 line, a bright component appeared with the similar velocity range to that of the blue hump in the H$\alpha$ line (spectrum H).
In the next four days, the bright blue shoulder grew wider and the H$\alpha$ line profile became like a top-hat profile (spectrum I).

By 16 September 2010, two orbital cycles after the giant outburst, both lines have significantly weakened (spectrum J). 

The equivalent widths (EW) of the H$\alpha$ and He I $\lambda$5876 lines around the giant outburst were highest in the last five years; EW(H$\alpha$) $\sim -18 {\rm \AA}$ and EW(He I $\lambda$5876) $\sim -1.5 {\rm \AA}$ [see figure \ref{LC_X} for EW(H$\alpha$), where the error size is denoted by the size of each symbol]. 
EW(H$\alpha$) showed complicated variability during our monitoring period: the absolute value of  EW(H$\alpha$), $|$EW(H$\alpha$)$|$, monotonically increased for about one orbital period until it temporarily decreased when the X-ray precursor occurred in October 2009 ($\sim$ JD 2455140). 
Afterwards, it increased for a while until two days before the peak of the outburst ($\sim$ JD 2455170). 
A brief low state in the H$\alpha$ line flux was seen for $\sim 10$ days between JD 2455169 and JD 2455178 (3 -- 12 December 2009), which was followed by a gradual decrease of $|$EW(H$\alpha$)$|$ until the end of the giant outburst.
Then, after a rapid increase two weeks prior to the precursor of the normal outburst in March 2010 ($\sim$ JD 2455270), it entered a long-term, declining phase.
Since July 2010, $|$EW(H$\alpha$)$|$ is at a level comparable to that in 2007 (\cite{Mor10}, $\sim$ $-$9 \AA).
\citet{Cam11b} reported a similar trend of variability of $|$EW(H$\alpha$)$|$ in their figure 1, although our $|$EW(H$\alpha$)$|$ is approximately 5 \AA \ smaller than theirs.

\section{Discussion}
The H$\alpha$ line profiles has exhibited remarkable variability not only during the 2009 giant outburst but also before and after it. 
The E/C ratio $\gtrsim 3.5$ during the giant outburst and one cycle before/after it (from August 2009 through March 2010) is significantly higher than the previous values ($\le 2.7$) since 2005.
Besides, since August 2009, after the bright normal outburst, the V/R variability has not followed any longer the 500-day periodic variation found by M10; the expected V/R ratio in this period was $>$ 1, while the obtained line profiles exhibited V $<$ R.
These facts indicate that not only the Be disk has grown denser, but also a non-axisymmetric bright region has appeared in the disk.
\citet{Neg98} reported that the process that took place at the same time as the 1994 giant outburst resulted in a change of both the period and the phase of V/R variability, which implied a major perturbation of the physical conditions in the Be disk.
\citet{Hai04} carried out long-term observations including the 1994 giant outburst and found that the Be disk began changing its structure one year before the giant outburst (see their figure 2).
Our observations suggest that, for the 2009 giant outburst, such a profound perturbation in the Be disk began about one cycle before the giant outburst. 
Further observations are needed to check whether the 500-day period is temporarily hidden and will show up again, or will be replaced by a different period, as seen at the time of the 1994 giant outburst.

The enhanced H$\alpha$ profile has a blue hump and a triple peak.
Although both spectral components were seen in the same period, from one cycle before the giant outburst to one cycle after it, only the blue hump exhibited variability that modulated with the orbital phase.
The particularly interesting feature of this variability was the appearance of a bright ``shoulder'' after periastron passage, $\phi_X$ $=$ 0.1 -- 0.15 (spectra C, H and I). 
Besides, the radial velocity of the blue hump, $\sim$ $-100$ -- 0 km s$^{-1}$, is similar to the velocity of the violet peak of double-peaked profiles that showed the V/R variability reported by M10. 
These features imply that the blue hump arises from the outermost region of the Be disk, where the density is higher than usual, possibly due to a one-armed density wave. 
Given the orbital elements of A0535+262 \citep{Fin94}, this part of the Be disk happens to be located near the periastron. 
In such a situation, when the neutron star passes the periastron, an enhanced mass transfer should take place from the dense part of the Be disk to the neutron star. 
With the observer's direction derived by \citet{Fin94}, gas streaming motion associated with this mass transfer also has negative line-of-sight velocities. 
It is, therefore, likely that the blue shoulder is the observational manifestation of the dense gas stream from the Be disk toward the neutron star.

Some Be stars also have shown triple-peaked profiles: e.g., $\phi$ Per and 59 Cyg. 
\citet{Mai04} pointed out the possibility that the triple-peaked profiles of these stars are caused by the small absorption by the companion star.
A0535+262 is, however, the binary of a Be star and a neutron star, and the dip between adjacent two peaks is much broader and deeper than that discussed in \citet{Mai04}.
It is therefore unlikely that the neutron star is involved in causing the triple-peaked profile.

\citet{Neg01b} observed long-term variation of the H$\alpha$ line of another Be/X-ray binary 4U0115+63, and found that the profile often shows a succession of single-peaked and shell profiles around a giant outburst (see their figures 2 and 6). 
Since the single-peaked and shell profiles are typical profiles when a Be disk is seen pole-on and equator-on, respectively, they interpreted this phenomenon as evidence for a precessing, warped Be disk driven by radiation from the central star\citep{Por98}. 
If the red triple-peaked profile of A0535+262 is due to a warping of the Be disk, the estimated time-scale of the precession is $\sim 30$ days, using table 1 and equation (8) of \citet{Por98}.
This time-scale is comparable with the time-scale of observed profile variability around the 2009 giant outburst. 
Although it is difficult to predict when and how the mass is transferred from the warped disk to the neutron star, it seems plausible that such a strong disturbance in the Be disk can shift the phase of maximum mass transfer toward the neutron star, and in turn the phase of X-ray maximum. 

The equivalent width of the H$\alpha$ line reflects the density in the outer part of the Be disk, since, in Be/X-ray binaries, the size of the Be disk, which is determined by the tidal truncation by the neutron star, does not change so much.
The gradual increase of $|$EW(H$\alpha$)$|$ and/or E/C, from 2005 through the 2009 giant outburst (figure \ref{LC_X}, see also M10), therefore, indicates that the Be disk grew denser and denser in the outer part.
Given this long-term trend, the temporary decrease of $|$EW(H$\alpha$)$|$ after each of the precursor and the giant outburst implies that a significant amount of mass in the disk was transferred to the neutron star during these events and that the Be disk recovered shortly after each event. 
The optical-IR photometry including the giant outburst in 1994 shows the same tendency \citep{Hai04}.

The H$\alpha$ line and He I $\lambda$5876 has shown qualitatively similar variations in the long term until August 2010.
After that, however, the He I $\lambda$5876 line profile became quite different from that of the H$\alpha$ line, e.g., significant decrease in the flux and the opposite trend in the V/R ratio.
Since the region emitting the He I line is slightly smaller than that emitting the H$\alpha$ line, it implies that the density and/or the temperature of the He I $\lambda$5876 emitting region decreased after a couple of mass transfer events.

\section{Concluding Remarks}
We observed the Be/X-ray binary A0535+262, covering the whole course of the 2009 giant outburst.
Observed emission line profiles during the giant outburst imply active components in the Be disk, which cause significant variability in the observation period.
The bright blue shoulder indicates the enhanced gas stream from the outermost part of the Be disk to the neutron star at periastron.
Highly redshifted, enhanced triple-peaked feature in the H$\alpha$ line profiles is possibly from a warping component. 
Further study is needed to clarify the nature of these features.

After the 2009 giant outburst, three normal outbursts occurred in A0535+262, followed by the next giant outburst in February 2011 \citep{Cam11}. 
As of March 2011, the Be disk of A0535+262 is still active according to its EW, so that monitoring observations covering several more cycles is important to study in more detail the interaction of the Be disk and the neutron star. 

\bigskip
We are very grateful to Bun'ei Sato for kindly observing A0535+262, and Tatehiro Mihara for constructive discussion.
This paper is based on observations taken at the Okayama Astrophysical Observatory and Gunma Astronomical Observatory.
This work was supported by Research Fellowships for the Promotion of Science for Young Scientists (YM).
We also acknowledge support of JSPJ Grant-in-Aid for Scientific Research (2054024, 18340055), the Grant-in-Aid for the Global COE Program "The Next Generation of Physics, Spun from Universality and Emergence" from the Ministry of Education, Culture, Sports, Science and Technology (MEXT) of Japan, and the Collaborative Research Program 2010, Information Initiative Center, Hokkaido University.



\begin{thebibliography}{}
\bibitem[Carciofi(2010)]{Car10}
	Carciofi, A.C. 2010, astro-ph, arXiv:1009.3969
\bibitem[Coe et al.(1975)]{Coe75}
	Coe M. .J., Carpenter G. F., Engel A. R. \& Quenby J. J. 1975, \nat, 256, 630
\bibitem[Camero-Arranz et al.(2011a)]{Cam11}
   Camero-Arranz A., A. Finger M. H., Wilson-Hodge C., Jenke P. A., \& Steele I. 2011a, Atel. \#3165
\bibitem[Camero-Arranz et al.(2011b)]{Cam11b}
   Camero-Arranz A et al 2011b, astro-ph, arXiv:1105.2081
\bibitem[Finger et al.(1994)]{Fin94}
	Finger M. H., Cominsky L. R., Wilson R. B., Harmon B. A. \& Fishman G. J. 1994, AIPC, 308, 459
\bibitem[Giangrande et al.(1980)]{Gia80} 
	Giangrande A., Giovannelli F.,Bartolini C., Guarnieri A. \& Piccioni A. 1980, \aaps, 40, 289
\bibitem[Haigh et al.(2004)]{Hai04}
   Haigh N. J., Coe M. J. and Fabregat J. 2004, \mnras, 350, 1457
\bibitem[Krimm et al.(2009)]{Kri09}
   Krimm H. A. et al. 2009, Atel. \#2336
\bibitem[Lee et al.(1991)]{Lee91}
   Lee, U., Saio, H., \& Osaki, Y. 1991, \mnras, 250, 432
\bibitem[Maintz et al.(2004)]{Mai04}
	Maintz M., Rivinius T., \v{S}tefl S., \& Stahl O. 2004, IAU Symp., 215, 95
\bibitem[M10]{Mor10}
	Moritani Y. et al. 2010, \mnras, 405, 467
\bibitem[Negueruela et al.(1998)]{Neg98}
	Negueruela I., Reig P., Coe M. J. \& Fabregat J. 1998, \aap, 336, 251
\bibitem[Negueruela \& Okazaki(2001)]{Neg01a}
	Negueruela I. \& Okazaki A. T. 2001, \aap, 369, 108
\bibitem[Negueruela et al.(2001)]{Neg01b}
	 Negueruela I., Okazaki A. T., Fabregat J., Coe M. J., Munari U. \& Tomov T. 2001, \aap, 369, 117
\bibitem[Okazaki \& Negueruela(2001)]{Oka01b}
	Okazaki A. T., Negueruela I. 2001, \aap, 377, 161  
\bibitem[Porter(1998)]{Por98}
	Porter J. 1998, \aap, 336, 966
\bibitem[Reig et al.(2007)]{Rei07}
	Reig P., Larionov V., Negueruela I., Arkharov A. A. \& Kudryavtseva, N. A. 2007, \aap, 462, 1081
\bibitem[Rosenberg et al.(1975)]{Ros75}
	Rosenberg F. D., Eyles C. J., Skinner G. K. \& Willmore A. P. 1975, \nat, 256, 628
\bibitem[Sugizaki et al.(2009)]{Sug09}
   Sugizaki M. et al. 2009, Atel. \#2277
\bibitem[Wilson et al.(2002)]{Wil02}
   Wilson C. A., Finger M. H., Coe, M. J., Laycock S. \& Fabregat, J. 2002, \apj, 570, 287
\bibitem[Wilson et al.(2008)]{Wil08}
   Wilson C. A. Finger M. H. \& Camero-Arranz A. 2008, \apj, 678, 1263
\bibitem[Wilson-Hodge et al.(2009)]{Wil09}
   Wilson-Hodge C. A., Finger M. H., Camero-Arranz A. \& Connaughton V. 2009, Atel. \#2324
\end{thebibliography}
\end{document}